\documentclass[
    ,final            
  ]
  {aipproc}

\layoutstyle{8x11double}

\begin{document}

\title{The use of Planetary Nebulae precursors in the study of Diffuse
Interstellar Bands}


\author{P. Garc\'\i a-Lario}{
  address={Research and Scientific Support Department of ESA. European
Space Astronomy Centre, Villafranca del Castillo, Madrid, Spain}
}

\author{R. Luna}{
  address={Escuela Polit\'ecnica Superior de Alcoy, Universidad Polit\'ecnica
  de Valencia, Spain}
}

\author{M.A. Satorre}{
  address={Escuela Polit\'ecnica Superior de Alcoy, Universidad Polit\'ecnica
  de Valencia, Spain}}

\begin{abstract}
We present the first results of a systematic search for Diffuse 
Interstellar Bands in a carefully selected sample of post-AGB stars 
observed with high resolution optical spectroscopy.
These stars are shown to be ideal targets to study this old, intriguing
astrophysical problem. Our results suggest that the carrier(s) of these
bands may not be present in the circumstellar environments of these evolved 
stars. The implications of the results obtained 
on the identification of the still unknown carrier(s) are discussed. 
\end{abstract}

\maketitle


\section{INTRODUCTION}

 Diffuse Interstellar Bands (DIBs, hereafter) are 
bands of variable strength and width of
still unknown origin which appear overimposed on the spectra of bright but 
heavily reddened stars. Since their discovery, now more than 80 years ago, they
have been associated to the interstellar medium, because their strength 
appear to be well correlated with the observed extinction.

 Currently, nearly 300 DIBs are catalogued (Galazutdinov et al. 2000) 
extending from ultraviolet to near-infrared wavelengths (3600--12000 \AA), 
the most studied ones being those found at 4430 \AA, 5780 \AA, 5797 \AA~ and 
6284 \AA. 

 Many carriers have been proposed, but for none of them convincing arguments
exist so far (for a review see Fulara \& Krelowski 2000). 
The problem still constitutes a major challenge for spectroscopists, 
astronomers and physicists.

 The suspicion is that the carrier(s) may have a molecular nature, as 
substructures ressembling the PQR-branches of ro-vibrational transitions
have been detected in some of these bands, when observed with very high
spectral resolution (Kerr et al. 1998). 
 
 On the other hand, the existence of \textit{families} of DIBs (Krelowski 
\& Walker 1987) suggests
that there is not a unique carrier. Moreover, the species implicated must be 
ubiquitous in space, since DIBs have been detected towards a wide variety of
astronomical sources, from comets to external galaxies. 

 The most promising hypothesis are: (i) long carbon chains, like 
polyacetylenes (Douglas 1977); (ii) PAH cations (Allamandola et al. 1999; 
Salama et al. 1999); or (iii) fullerenes (Foing \& Ehrenfreund 1994).
 
 There are strong evidences that the relative strength of
DIBs are correlated with the properties of the clouds in the line of sight.
This environmental dependence may reflect an interplay of ionization, 
recombination, dehydrogenation and destruction of chemically stable, 
carbonaceous species (Salama et al. 1996).

 Investigations of DIBs in regions of different metallicity, chemical 
properties and UV radiation field may allow us to constrain the 
physico-chemical properties of the (different) DIB carriers.

 Unfortunately, it is difficult to probe the interstellar medium along a 
given line of sight; usually this is a combination of many different 
clouds with inhomogeneous properties and complex morphologies.

\section{Diffuse \textit{Circumstellar} Bands}

 Since circumstellar shells are sources of replenishment of the interstellar 
medium, it seems reasonable to expect that DIB carriers may also be present 
in some of these shells. In particular, the suspected connection between DIB 
carriers and some carbon-rich compounds can be investigated attending to the 
usually known chemistry and physical properties of these circumstellar shells. 

  A first attempt to detect Diffuse \textit{Circumstellar} Bands was carried
out by Le Bertre \& Lequeux (1993) using a sample containing a mixture of 
carbon-, oxygen- and nytrogen-rich mass-losing stars. However, they 
failed to reach any firm conclusion from the results obtained. 
In particular, they did not detect any band 
in the spectra of sources with strong PAH emission at mid-infrared wavelengths,
contrary to their expectations. In contrast, strong DIBs were observed toward
other carbon-rich sources, as well as toward most of their oxygen-rich and 
nytrogen-rich sources in the sample.

 Observationally, the detection of Diffuse Bands (DBs, hereafter) around
evolved stars is hampered by the fact that most mass-losing stars are usually 
strongly variable stars, 
surrounded by very cool extended atmospheres where molecules are the dominant 
source of opacity. These stars are very difficult to model and DBs are
hardly detected against the forest of features attributed to 
molecular transitions which appear overimposed on the stellar continuum.
This has prevented the systematic search for DBs in evolved stars in the 
past.

\section{POST-AGB STARS AS IDEAL TARGETS FOR DB STUDIES}

  Fortunately, an alternative exists which have so far not yet been 
explored. These are the post-AGB stars, rapidly evolving stars in the
transition from the Asymptotic Giant Branch (AGB) to the Planetary Nebula (PN)
stage. While in the early post-AGB stage, these stars are still surrounded by 
the relatively thick circumstellar shells formed during the 
previous mass-losing AGB phase. Their central stars show a wide variety of 
spectral types ranging from M to B in which seems to be an evolutionary 
sequence in their way to become PNe. And the chemical composition of
the gas and dust in the shell can easily be determined from observations at
optical, infrared, mm/submm or radio wavelengths. In addition, they are
located in many cases at relatively high galactic latitudes, which favours 
the potential circumstellar origin of the features observed. 

 In order to make a systematic study of the presence of these bands in 
post-AGB stars, we carefully selected a sample of 33 sources displaying 
all kind of spectral types from G to B\footnote{ Stars with spectral types 
later than G were discarded for the analysis, to
facilitate the identification of the features against the stellar continuum.}.
and covering a wide range of galactic latitudes. It contains a mixture of 
C-rich and O-rich stars with  a well determined 
value of the colour excess E(B$-$V), as a reddening indicator, .

 The sample was then split in two subgroups according to whether the overall
extinction observed is predominantly interstellar or circumstellar in origin
(see the details of how this classification was done in Luna et al. 2005; in
preparation).

  High resolution spectra taken at various telescopes covering the spectral
range 4000 -- 10000 \AA~ (many of them originally taken for chemical 
abundance analysis purposes at the VLT and kindly provided by Hans van 
Winckel and collaborators) were used for our analysis.

 \section{RESULTS}

 Up to nine different DBs were investigated in detail. Table 1 lists the
wavelengths corresponding to these features; their typical FWHM, as a way 
to characterize their broadness; and their 
sensitivity to the reddening, measured as EW/E(B$-$V), re-derived by us 
from the analysis of a sample of field stars compiled by Thornburn et al. 
(2003) for six of these DBs and by Jenniskens \& D\'esert (1994) for three
other ones.


\begin{table}
\begin{tabular}{ccc}
\hline
   \tablehead{1}{c}{b}{DB} &
   \tablehead{1}{c}{b}{FWHM} &
   \tablehead{1}{c}{b}{EW/E(B$-$V)}   \\
(\AA)& (\AA)& (\AA)/mag \\
\hline
5780 & 2.2 & 0.46  \\ 
5797 & 1.1 & 0.19  \\
5850 & 1.1 & 0.050 \\
6196 & 0.9 & 0.053 \\
6284 & 4.5 & 1.05  \\
6379 & 1.1 & 0.093 \\
6614 & 1.2 & 0.21  \\
6993 & 1.6 & 0.12  \\
7224 & 1.3 & 0.25  \\
\hline
\end{tabular}
\caption{Main characteristics of the Diffuse Bands included in our
 analysis}
\label{tab:a}
\end{table}

 Note that to distinguish these weak features from weak stellar
lines or telluric contaminations is not always a simple task and makes it
necessary to use detailed stellar models (to subtract the atmospheric
features) and high resolution spectroscopy (to properly remove undesired
contaminations), as the only way to derive the accurate strength of each band.

 In Figure 1 we show the results obtained for the 6284 \AA~
feature, which is not only the strongest but also the broadest band 
included in our analysis and, as such, relatively easy to measure, in spite
of the contamination by telluric lines, which must be carefully removed.  

\begin{figure}
  \includegraphics[height=.5\textheight]{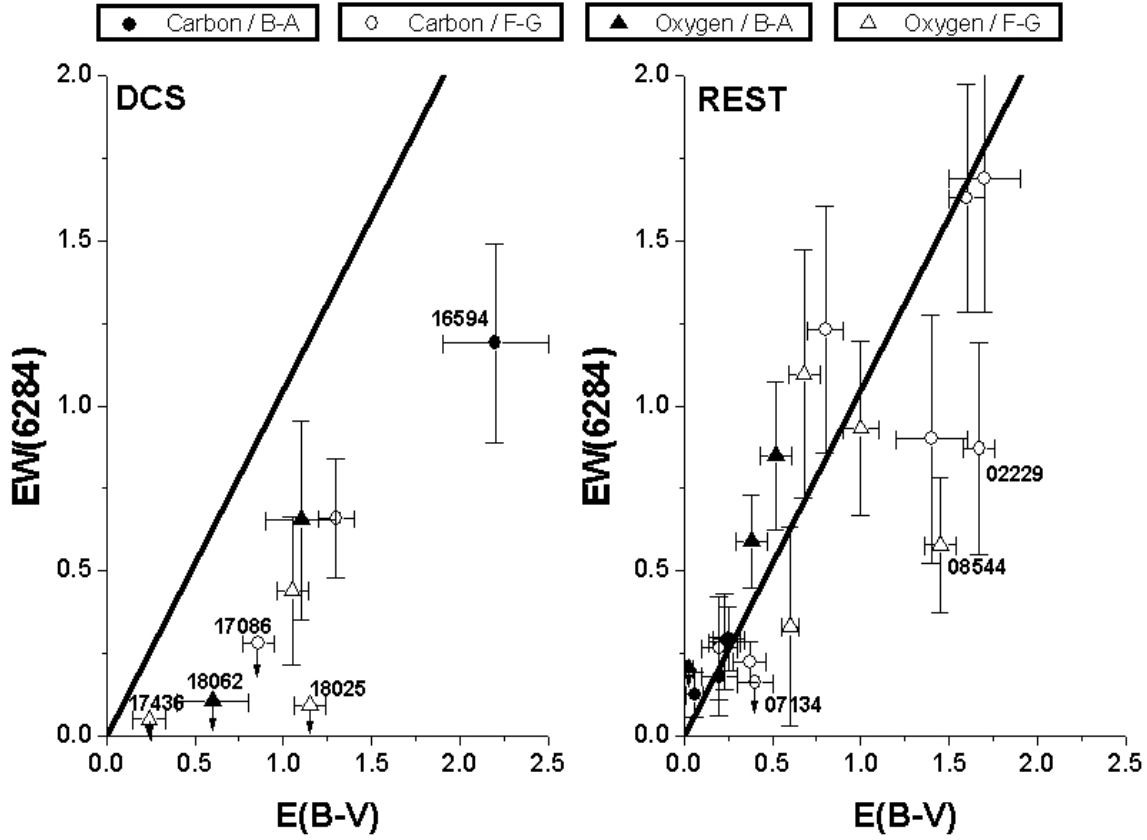}
  \caption{Equivalent width of the 6284 \AA~ feature as a function of the
total reddening observed in the direction of the sources in our sample. 
Sources dominated by circumstellar extinction (left) are compared to
those for which intestellar extinction seems to be the dominant 
contributor to the observed reddening. The solid line represents the
correlation found in field stars, and is shown for comparison. Different
symbols are used for C-rich and O-rich stars and for early-type (B-A) stars
and intermediate-late (F-G) type stars.}
\end{figure}

 As we can see, the strength of the 6284 \AA~ feature in sources 
dominated by interstellar extinction follows the same trend observed
in field stars, supporting that there is a tight correlation between the
band strength and interstellar reddening. In contrast, all sources which 
appear to be dominated by circumstellar extinction, show a DB strength 
far below the value expected for the observed reddening. In the most 
extreme cases, we find stars in which the 6284 \AA~ band is not even 
detected, while they are considerably reddened in the optical. 

 These results are applicable to all stars in the sample, independent on
whether they are C-rich or O-rich, and/or whether they are early-type
stars (B-A) or intermediate-late (F-G) type stars.  A few outliers observed
in the right panel, like IRAS 02229+6208 or IRAS 08544-4431, with labels in
the figure, are also observed as outliers when other DBs are analysed.  
These stars may have been incorrectly identified as dominated by interstellar
extinction.

 The same trend is observed when other DBs are investigated. Figure 2
shows the results obtained with the 7224 \AA~ feature, as an example. 
In general, the strengths of the DBs are found to be well correlated with the
interstellar extinction only in those sources showing little circumstellar 
contribution to the overall reddening, while DBs are weak or absent in
sources dominated by circumstellar reddening.


\begin{figure}
  \includegraphics[height=.5\textheight]{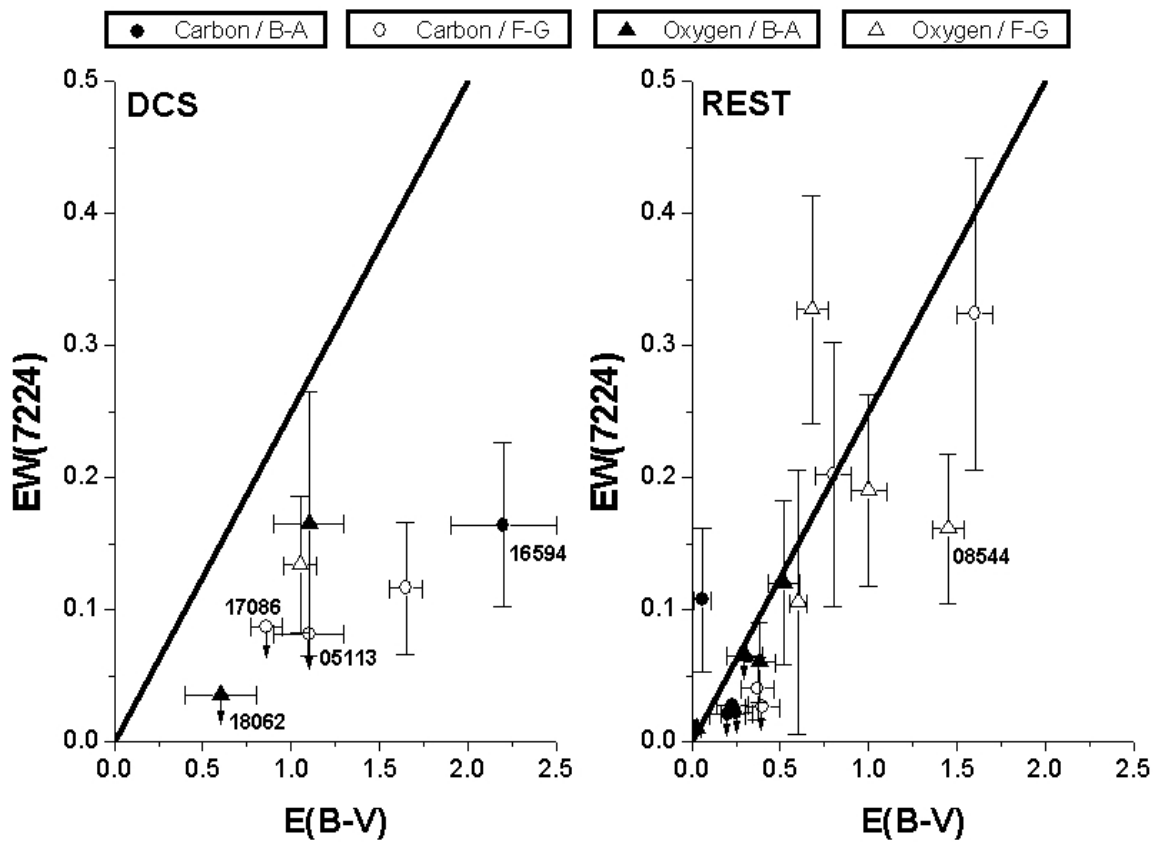}
  \caption{Same as Fig. 1, for the 7224 \AA~ feature.}
\end{figure}


\section{DISCUSSION}

 Globally considered our results suggest that the carrier(s) of DBs must not
be present in the circumstellar shells of post-AGB stars. At least, 
not under the environmental conditions needed to excite the transitions 
which we identify as DBs in the interstellar medium.

  In particular, we do not find any evidence of the carbonaceous nature 
of the carrier(s), something generally accepted in the literature, nor any 
correlation with the presence of PAHs in the mid-infrared spectrum of these 
sources, as it has been claimed by several authors in the past. 

 If DBs are connected with PAHs or with any other carbonaceous species such
as the ones suggested in the introduction of this paper, 
their carrier(s) must form at a later stage, under different conditions, when
the envelope of the post-AGB star is totally diluted in the interstellar
medium as a result of the expansion of the shell, perhaps as a result of
their processing by the hard UV photons present in the interstellar medium.

 In this sense, the identification of the carriers as strongly ionized PAHs
and/or radicals liberated from carbonaceous species as a consequence of 
photoevaporation of dust grains in the interstellar medium  would be 
consistent with our observations.   

  In order to confirm this hypothesis we plan to extend our analysis in the
near future to other sources in a similar evolutionary stage and study 
in particular very young C-rich PNe in which the UV field may be already 
strong enough to produce the \textit{in situ} formation of some of these
carriers. 

\begin{theacknowledgments}
 This work was partially funded by grants AYA2003--09499 and AYA2004--05382 of 
the Spanish Ministerio de Ciencia y Tecnolog\'\i a.
\end{theacknowledgments}



\bibliographystyle{aipproc}   



\end{document}